\documentclass[11pt]{article}
\usepackage{amsmath,amssymb,graphics}
\usepackage{bbold}
\usepackage{amsthm}
\usepackage{graphicx}
\usepackage{geometry}
\usepackage[all]{xy}
\usepackage{url}
\usepackage{setspace}
\usepackage[square ,numbers, sort&compress, merge]{natbib} 
\usepackage{fullpage}
\usepackage[linktocpage=false,colorlinks,pdfborder={0 0 0},citecolor=black,urlcolor=black,pdfstartview={FitH},linkcolor=black]{hyperref}

\newcommand{\ket}[1]{|#1\rangle}
\newcommand{\bra}[1]{\langle #1|}
\newcommand{\braket}[2]{\langle #1|#2\rangle}

\theoremstyle{plain}

\theoremstyle{definition}

\theoremstyle{remark}

\numberwithin{equation}{section}

\title{On the Symplectic Reduced Space of Three-Qubit Pure States}

\author{Saeid Molladavoudi}

\date{\vspace{-5ex}}

\begin{document}

\maketitle

\begin{abstract} Given a specific spectra of the single-particle reduced density matrices of three qubits, the singular symplectic reduction method is applied to the projective Hilbert space of tripartite pure states, under the local unitary group action. The symplectic structure on the principal stratum of the symplectic quotient is obtained. A criterion from which the elements of the local normal model of the principal stratum can be constructed up to an equivalence relation and also the components of the reduced Hamiltonian dynamics on it are investigated. It is discussed that other lower dimensional strata are isolated points and so they are the fixed points of every reduced Hamiltonian flow, i.e. relative equilibria on the original manifold. \end{abstract}


\section{Introduction} \label{introduction}

The geometry and topology of the space of entanglement types, or the \textit{orbit space}, of a composite quantum system under the Local Unitary Group $K$-action can be studied using the algebra of the $K$-invariant polynomials \cite{linden1998}, since the local unitary group $K = SU(n_j)^{\times_N}$ is a compact Lie group acting on the projective Hilbert space $M = \mathcal{P}(\mathcal{H})$ of the multipartite pure states as a compact K\"ahler manifold, with $\mathcal{H} = \otimes_{j=1}^{N}\mathcal{H}_j$. Therefore, the $K$-action on the K\"ahler manifold $M$ is proper and the orbit space $M/K$ is a Hausdorff space. A Hilbert's theorem ensures that the algebra of $K$-invariant polynomials are finitely generated \cite[Section 8.14]{weyl1997} and so the orbit space $M/K$ is locally homeomorphic to the image of the corresponding Hilbert map, as a semi-algebraic  subset of $\mathbb{R}^d$ given by polynomial equalities and inequalities, where $d$ is the number of the basis of the algebra of $K$-invariant polynomials \cite{friedman1963}.

In general, if $M$ is a compact symplectic manifold, equipped with a closed non-degenerate $2$-form $\omega$, under the proper and Hamiltonian action of the symmetry Lie group $K$, the orbit space $M/K$ is a stratified Poisson space, such that the strata are Poisson manifolds and the Poisson subalgebra $C^{\infty}(M)^K$ of $K$-invariant smooth functions can separate the $K$-orbits in $M$ \cite{arms1991,cushman2001}. Alternative approach exists in the context of symmetry reduction of Hamiltonian systems, in which the components of the moment map $\textbf{J}:M \rightarrow \mathfrak{k}^*$, where $\mathfrak{k}^*$ is the dual of the Lie algebra $\mathfrak{k}$ of $K$, are conserved with respect to the integral curves of the Hamiltonian vector fields, i.e. the \textit{N\"oether's theorem}. It is initiated with the Marsden-Weinstein regular reduction in \cite{weinstein1971} and continued in \cite{sjamaar1991,bates1997,ortega2004} for the singular reduction of Hamiltonian manifolds. In the latter case, the resulting symplectic quotient $M_{\xi} = \textbf{J}^{-1}(\xi)/K_{\xi}$ is a stratified symplectic space, in which the strata are symplectic manifolds. The advantage is that this method enables us to obtain the components of the reduced dynamics on the symplectic strata $M_{\xi}^{(H)}$ of the reduced space $M_{\xi}$.

In the current paper, we consider the complex projective Hilbert space $\mathcal{P}(\mathcal{H})$ of a tripartite pure states, as a K\"ahler manifold, which is acted upon properly and in a Hamiltonian fashion by the local unitary group $K = SU(2)^{\times_3}$, and the corresponding symmetry or the conservation law is the preservation of the (shifted) spectra of the single-particle reduced density matrices, encoded in the components of the associated moment map \cite{sawicki2011a,sawicki2011b}. However, this action is not free and so the values $\xi \in \mathfrak{k}^*$ are \textit{singular} values of the associated moment map $\textbf{J}$, since for some points $p \in M$, the corresponding isotropy subgroup $K_p \subset K$ is continuous, i.e. for each $\xi \in \mathfrak{k}^*$, $\textbf{J}^{-1}(\xi)$ contains a point $p \in M$, such that the corresponding isotropy subgroup $K_p$ is continuous. Therefore, we consider the symplectic singular reduction method in order to obtain the reduced symplectic quotient $M_{\xi} = \textbf{J}^{-1}(\xi)/K_{\xi}$, for a given $\xi \in \mathfrak{k}^*$, with respect to the conservation of the single-particle reduced density matrices. The geometry and topology of the $K$-orbit space of tripartite pure states for three qubits are studied in \cite{walck2005}, using the algebra of $K$-invariant polynomials, and in \cite{iwai2007b}, using the bipartite decomposition procedure. 

In \cite{sawicki2012b}, the dimension of the principal stratum of the symplectic quotient $M_{\xi}$, for a fixed $\xi$ in the associated moment polytope, is studied in the case of tripartite pure states of three qubits. In the current paper, which is motivated by the work in \cite{sawicki2012b}, by Sawicki \textit{et al}, the symplectic structure $\omega^{(\text{prin})}_{\xi}$ on the principal stratum $M^{(\text{prin})}_{\xi}$ of the stratified symplectic space $M_{\xi}$ is derived for a fixed $\xi \in \text{int}(\Delta)$, where $\Delta$ represents the moment polytope of the proper Hamiltonian action of $K$ on $M = \mathcal{P}(\mathcal{H})$. In addition, a criterion is obtained from which the elements of the local normal form on the principal stratum $M^{(\text{prin})}_{\xi}$ can be determined up to the free action of the principal isotropy subgroup. Moreover, from the symplectic structure, the components of the reduced Hamiltonian dynamics, namely the reduced Hamiltonian functions $f^{(\text{prin})}_{\xi} \in C^{\infty}(M^{(\text{prin})}_{\xi})$ as well as their induced Hamiltonian vector fields, are investigated on the principal stratum. Furthermore, by definition of a stratified symplectic space, the symplectic structure on all lower dimensional strata can be obtained by the symplectic structure on the principal stratum. From the quantum mechanical point of view, the reduced dynamics on the principal stratum can shed some light on the non-local perturbations of generic points of tripartite pure states of three  qubits, whose entanglement is invariant under the local unitary operations. 

The outline of the paper is as follows. In section \ref{review-of-symplectic-singular-reduction}, the singular symplectic reduction is briefly reviewed. In section \ref{review-of-local-unitary-action}, the local unitary action on the projective Hilbert space of a composite quantum system, as well as the preliminary notation, is introduced. In section \ref{singular-reduction-of-local-unitary-action}, the singular symplectic reduction method is applied to the projective Hilbert space under the local unitary action and the local normal form, the symplectic structure on the principal stratum, the reduced dynamics on it and the dynamics on other lower dimensional strata are studied in further details. Finally in section \ref{conclusions} we summarize the results.
\section{Review of Symplectic Singular Reduction} \label{review-of-symplectic-singular-reduction}

Let $(M,\omega)$ be a connected symplectic manifold and $K$ a compact Lie group acting \textit{properly} on $M$. Recall that the action of the Lie group $K$, i.e. $\Phi: K \times M \rightarrow M$ and the Lie algebra $\mathfrak{k}$, i.e. $\phi : M \times \mathfrak{k} \rightarrow \mathfrak{X}(M)$ are \textit{continuous} and \textit{infinitesimal} symmetry actions respectively, where $\mathfrak{X}(M)$ denotes the Lie algebra of the smooth vector fields defined on $M$ and equipped with the Lie bracket. The Lie group action $K \ni g \mapsto \Phi_g \in \text{Diff}(M)$ is a group homomorphism, whereas the Lie algebra action $M \times \mathfrak{k} \ni (p,X) \mapsto \phi_X(p) \in TM$ is a Lie algebra anti-homomorphism. The fundamental vector fields of $X \in \mathfrak{k}$, or the \textbf{infinitesimal generators} $\phi_X \equiv X_M \in \mathfrak{X}(M)$, of the Lie group $K$-action is defined by 
\[
\phi_X(p) \equiv X_M(p) := \left. \frac{d}{dt} \right|_{t=0} \Phi_{\exp(tX)}(p), \quad p \in M,
\]
which constitutes the Lie algebra $\mathfrak{k}$-action. Given a point $p \in M$, the closed Lie subgroup $K_p := \left\{ \left. g \in K \right| \Phi_g(p) = p \right\}$ is called the \textit{stabilizer}, or the \textit{isotropy} subgroup of the point $p \in M$. Therefore, the Lie subalgebra $\mathfrak{k}_{p} := \left\{ \left. X \in \mathfrak{k} \right| \phi_X(p) = X_M(p) = 0 \right\}$ is called the stabilizer, or isotropy or symmetry subalgebra of $p \in M$. In fact, $\mathfrak{k}_p$ is the Lie algebra of the closed Lie subgroup $K_p$. Moreover, since for every $g \in K$ and $p \in M$, $K_{g.p} = g K_p g^{-1}$ the isotropy subgroups of the points in the same orbit are all conjugate in $K$. This conjugation establishes an equivalence relation $ \sim $ in the space of isotropy subgroups of the Lie group $K$, i.e. $K_p \sim K_q$ if and only if there exists a $g \in K$ such that $K_p = g K_q g^{-1}$. The equivalence class $(K_p) := \left\{ g K_p g^{-1}\right\}_{g \in K}$ is called the $(K_p)$-\textit{orbit type} of the orbits through $p \in M$. Equivalently, two orbits $K.p$ and $K.q$ have the same orbit type, if $K_p$ and $K_q$ are conjugate in $K$. Let $\mathcal{I}$ denotes the set of orbit types of the $K$-action on $M$, then we can introduce a partial ordering on $\mathcal{I}$ by the fact that $(K_q) \leq (K_p)$ if and only if $K_p$ is conjugate to some subgroup of $K_q$ in $K$. If the $K$-action $\Phi$ is proper, the orbit space $M/K$ is a Hausdorff topological space, which is the case for compact Lie groups, such as the local unitary group. The union of orbits having the same orbit type is called a orbit type stratum of $M$ and is denoted by $M_{(H)}$ through $p \in M$, such that $H$ is conjugate in $K$ with $K_p$, and its image under the projection $\pi:M \rightarrow M/K$ is called the orbit type stratum of $M/K$ and is denoted by $M_{(H)}/K$ through $\pi(p) \equiv x \in M/K$. 

Furthermore, if $K$ is a compact Lie group acting on a connected smooth manifold $M$, then there exists a unique minimal orbit type $(H_{\text{prin}})$, such that the stratum associated to the $(H_{\text{prin}})$-orbit type is connected, open and dense in $M/K$, for which the dimension equals to $\text{dim}(M) - \text{dim}(K) + \text{dim}(H_{\text{prin}})$ \cite[Theorem 1.4]{schwarz1980}. Such orbit type is called the \textit{principal orbit type} and the corresponding orbits are called principal orbits. In other words, an orbit $K.p$ is a principal orbit, if and only if for all $q \in M$, the isotropy subgroup $K_p$ is conjugate in $K$ to some subgroup of $K_q$. Also, the corresponding stratum $M_{(\text{prin})}/K$ in $M/K$ is called the principal stratum. The dimension of the principal stratum determines the dimension of the orbit space.

According to the classical \textbf{N\"oether's theorem}, for a symplectic manifold $(M,\omega)$ acted upon by the Lie group $K$ in a Hamiltonian fashion, the components of the associated equivariant moment map $\textbf{J} : M \rightarrow \mathfrak{k}^*$ are preserved during the Hamiltonian dynamics, i.e. $\textbf{J} \circ \varphi_t = \textbf{J}$, where $\varphi_t$ is the corresponding Hamiltonian flow. The symplectic manifold $(M,\omega)$, endowed with a symmetry or a conservation law, may be reduced to the corresponding quotient space $M_{\xi}$ containing the equivalence classes of the level set of the moment map $\textbf{J}^{-1}(\xi)$ under the action of the isotropy group $K_{\xi}$, for a fixed $\xi \in \mathfrak{k}^*$. If the action of the group $K$ on the manifold $M$ is \textit{free}, i.e. $K_p = \left\{ e \right\}$, for all $p \in M$, and assuming that the action of the closed subgroup $K_{\xi}$ is free and proper on $\textbf{J}^{-1}(\xi)$, then the resulting quotient space $M_{\xi} := \textbf{J}^{-1}(\xi) / K_{\xi}$, for a given \textit{regular} value of the moment map $\xi \in \mathfrak{k}^*$, would be another symplectic manifold equipped with the induced symplectic structure $\omega_{\xi}$, defined by $\pi^*_{\xi} \, \omega_{\xi} = i^*_{\xi} \, \omega$, where $\pi^*_{\xi} : \textbf{J}^{-1}(\xi) \rightarrow M_{\xi}$ is the projection map to the symplectic reduced space and $i^*_{\xi} : \textbf{J}^{-1}(\xi) \hookrightarrow M$ is the inclusion map. This reduction procedure is known as the \textbf{Meyer-Marsden-Weinstein} \citep{meyer1973,marsden1974}, or the \textbf{regular symplectic point reduction} \citep{ortega2004}, since the point $\xi \in \mathfrak{k}^*$ is fixed.

If the condition on the freeness of the Hamiltonian action of the Lie group $K$ on the symplectic manifold $(M,\omega)$ is dropped, then the level set $\textbf{J}^{-1}(\xi)$ is a topological space. The orbit space $M_{\xi} := \textbf{J}^{-1}(\xi) / K_{\xi}$ is endowed with the quotient topology. In \citep{sjamaar1991} it is shown that for the Hamiltonian action of a compact Lie group $K$, the quotient space $M_{\xi=0}$ is a stratified symplectic space, satisfying the Whitney's condition, i.e. each strata is a symplectic manifold and the pieces are glued together nicely. This result is extended to the case of the proper action of a Lie group $K$ in \citep{bates1997}. For more details one can refer to \citep{ortega2004}.

Now, let $(M,\omega,K, \textbf{J})$ be a Hamiltonian $K$-space, where $(M,\omega)$ is a symplectic manifold acted upon properly and symplectically by the compact Lie group $K$ and $\textbf{J}: M \rightarrow \mathfrak{k}^*$ is the associated equivariant moment map, such that $\textbf{J}(p) = \xi$, with $p \in M$ and $\xi \in \mathfrak{k}^*$ as a value of $\textbf{J}$. Let the isotropy subgroup $K_p$ is denoted by $H$. Then,
\begin{itemize}
\item $\textbf{J}^{-1}(\xi) \cap M_{(H)}$ is a submanifold of $M$.
\item The set $M^{(H)}_{\xi} := (\textbf{J}^{-1}(\xi) \cap M_{(H)}) / K_{\xi}$ has a unique quotient differentiable structure such that the projection map $\pi^{(H)}_{\xi} : \textbf{J}^{-1}(\xi) \cap M_{(H)} \rightarrow M^{(H)}_{\xi}$ is a surjective submersion.
\item $(M^{(H)}_{\xi}, \omega^{(H)}_{\xi})$ is a symplectic manifold, where the symplectic structure $\omega^{(H)}_{\xi}$ is defined by
\begin{equation}
(i^{(H)}_{\xi})^* \omega = (\pi^{(H)}_{\xi})^* \omega^{(H)}_{\xi},
\label{reduced-symplectic-structure-on-the-strata-eq}
\end{equation}
where $i^{(H)}_{\xi}: \textbf{J}^{-1}(\xi) \cap M_{(H)} \hookrightarrow M$ is the inclusion map. $(M^{(H)}_{\xi}, \omega^{(H)}_{\xi})$ is called the \textit{singular symplectic point stratum}.
\item The connected components of $\textbf{J}^{-1}(\xi) \cap M_{(H)}$ are left invariant under the flow $\varphi_t$ of the Hamiltonian vector field $X_{h}$, for $h \in C^{\infty}(M)^{K}$, which also commutes with the $K_{\xi}$-action. Therefore, the induced flow $\varphi_t^{(H)}$ on $M^{(H)}_{\xi}$ is defined by
\begin{equation}
\pi^{(H)}_{\xi} \, \circ \, \varphi_t \, \circ \, i^{(H)}_{\xi} = \varphi^{(H)}_{t} \, \circ \, \pi^{(H)}_{\xi}.
\label{reduced-flow-on-the-strata-eq}
\end{equation}
\item The reduced Hamiltonian function $h^{(H)}_{\xi} : M^{(H)}_{\xi} \rightarrow \mathbb{R}$ of the flow $\varphi^{(H)}_t$ on $M^{(H)}_{\xi}$ is defined by 
\begin{equation}
h^{(H)}_{\xi} \, \circ \, \pi^{(H)}_{\xi} = h \, \circ \, i^{(H)}_{\xi}.
\label{reduced-hamiltonian-functions-on-the-strata}
\end{equation}
\end{itemize}
Then the quotient space $M_{\xi}$ is a stratified symplectic space with $(M^{(H)}_{\xi}, \omega^{(H)}_{\xi})$ as the strata. This is called the \textit{symplectic stratification theorem}. For more details and the proofs one can refer to \cite{sjamaar1991,bates1997,ortega2004}. Also, for more details on the relation between the symplectic leaves of the orbit space $M/K$, as a stratified Poisson manifold for a proper action of the Lie group $K$, and the symplectic strata introduced above one can refer to \cite{cushman2001}.
\section{Review of Local Unitary Action} \label{review-of-local-unitary-action}

Let's consider a composite quantum system, consisting of $N$ distinguishable $n_i$-level quantum subsystems, for $i=1, \cdots, N$, in its global pure state. Using the language of geometric quantum mechanics, the space $\mathcal{P}(\mathcal{H})$ of global quantum pure states is a $(\prod_{i=1}^{N} \, n_i - 1)$-dimensional K\"ahler manifold, equipped with both Riemannian and Symplectic structures, induced from the real and imaginary parts of the Hermitian inner products in $\mathcal{H} = \bigotimes_{N} \mathcal{H}_i$ respectively. Moreover, since the projective manifold $M=\mathcal{P}(\mathcal{H})$ is equipped with the transitive action of the unitary Lie group $U(\mathcal{H})$, the infinitesimal generators $Y_M(p)$ span the tangent space $T_pM$, for all $Y \in \mathfrak{u}(\mathcal{H})$. So, let $A,B \in \mathfrak{u}^*({\mathcal{H}})$, be the observables acting linearly on $\mathcal{H}$. The symplectic structure $\omega$ at $ p \in \mathcal{P}(\mathcal{H})$ is defined by \cite{benvegnu2004}
\begin{equation}
\omega_{p}(X_M,Y_M) := \frac{\text{i}}{2} \frac{\braket{ \psi}{[A,B] \psi}}{\braket{\psi}{\psi}} = \frac{\text{i}}{2} \text{Tr}(\rho_{\psi}[A,B]),
\label{projective-symplectic-structure-eq1}
\end{equation}
where $X_M,Y_M \in T_p \mathcal{P}(\mathcal{H})$ are given by
\[
X_M(p) = \left. \frac{d}{dt} \right|_{t=0} \pi(\exp(-\text{i} At) \, \psi) = -\text{i}\, [A,\rho_{\psi}], \quad A \in \mathfrak{u}^*(\mathcal{H}),
\]
for all $p \in \mathcal{P}(\mathcal{H})$, where $\pi: \mathcal{H} \rightarrow \mathcal{P}(\mathcal{H}), \, \psi \mapsto p \equiv \rho_{\psi}$ is the canonical projection. In fact $p \in \mathcal{P}(\mathcal{H})$ is the global pure state of the composite quantum system, which is either a \textit{separable} or an \textit{entangled} state. Note that in this paper the points in the projective Hilbert space $M = \mathcal{P}(\mathcal{H})$ are denoted by $p$ and $\rho_{\psi}$ interchangeably.

The corresponding local unitary transformation Lie group $K$ is the compact Lie group $K= SU(n_i)^{\times_N}$ acting on the manifold $\mathcal{P}(\mathcal{H})$, where $\mathcal{H} = \bigotimes_{N} \mathcal{H}_i$ and $\mathcal{H}_i = \mathbb{C}^{n_i}$, for $i = 1 , \cdots , N$. The natural action of the group $K$ on $\mathcal{H}$, i.e. $g \, . \psi = g_1 \psi_1 \otimes \cdots \otimes g_N \psi_N \in \mathcal{H}$, for $g = (g_1, \cdots , g_N) \in K$ and $\psi = \psi_1 \otimes \cdots \otimes \psi_N \in \mathcal{H}$, with $\psi_i \in \mathcal{H}_i$, is then projected to $M = \mathcal{P}(\mathcal{H})$ to determine the action of $K$ on the K\"ahler manifold $M$, namely
\begin{equation}
\Phi: K \times M \rightarrow M, (g,p) \mapsto \Phi_g(p) = g \rho_{\psi} g^{-1},
\label{local-unitary-group-action-on-projective-hilbert-space-eq}
\end{equation}
where $p \equiv \rho_{\psi} = \ket{\psi}\bra{\psi}/\braket{\psi}{\psi} \in M \equiv \mathcal{P}(\mathcal{H})$. Therefore, $X_M(p) = \dot{\rho}_{\psi} = -\text{i} \, [\eta,\rho_{\psi}]$, where $\eta \in \mathfrak{k}^*$. The isomorphism $\mathfrak{k} \cong \mathfrak{k}^*$, as well as for $\mathfrak{u}(\mathcal{H}) \cong \mathfrak{u}^*(\mathcal{H})$, is due to the Killing-Cartan metric defined on $K$, i.e. $\langle X , Y 
\rangle:= - \text{Tr}(XY)/2$, for $X,Y \in \mathfrak{k}$, i.e. if $\eta \in \mathfrak{k}^*$, then $-\text{i} \, \eta = X \in \mathfrak{k}$ and the Killing-Cartan metric is satisfied for an arbitrary $Y \in \mathfrak{k}$. The action of the complex structure $J$ on $X_M(p)$ reads as $JX_M(p) = \text{i} \, \dot{\rho}_{\psi} = [\eta,\rho_{\psi}]$, for $\eta \in \mathfrak{k}^*$.

The action of the local unitary group $K$ is \textit{proper} and \textit{symplectic}, since the Lie group $K$ is a compact Lie group preserving the symplectic structure of the K\"ahler manifold $\mathcal{P}(\mathcal{H})$, i.e. $\Phi^*_g \omega = \omega$, for every $g \in K$. Furthermore, this action is Hamiltonian and so there exists an equivariant moment map $\textbf{J}:M \rightarrow \mathfrak{k}^*$, defined by \cite{benvegnu2004}
\begin{equation}
\langle \textbf{J}(p) , X \rangle = \frac{\text{i}}{2} \frac{\braket{\psi}{X \, \psi}}{\braket{\psi}{\psi}} = \frac{\text{i}}{2} \text{Tr}(X \rho_{\psi}) \equiv J_X(p), \quad X \in \mathfrak{k},
\label{equivariant-moment-map-eq}
\end{equation}
where $J_X: M \rightarrow \mathbb{R}$ is the corresponding Hamiltonian function. The above-mentioned local unitary group $K$ is a subgroup of the unitary transformations of the global Hilbert space, i.e. $U(\mathcal{H})$, with dual of the Lie algebra $\mathfrak{k}^* = \mathfrak{su}^*(n_1) \oplus \cdots \oplus \mathfrak{su}^*(n_N)$. Hence, the quadruple $(M,\omega,K,\textbf{J})$ is a Hamiltonian $K$-manifold, with the moment map $\textbf{J}:M \rightarrow \mathfrak{k}^*$, given by \cite{sawicki2011a,sawicki2011b}
\begin{equation}
\textbf{J}(p) = (\rho^{(1)}-\frac{1}{n_1}\mathbb{1}_{n_1}) \oplus (\rho^{(2)}-\frac{1}{n_2}\mathbb{1}_{n_2}) \oplus \cdots \oplus (\rho^{(N)}-\frac{1}{n_N}\mathbb{1}_{n_N}) \in \mathfrak{k}^*,
\label{moment-map-eq}
\end{equation}
where $\rho^{(j)}$, for $j = 1 , \cdots,N$, represents the $j$th-subsystem's reduced density matrix, and the shifting $\rho^{(j)}-\frac{1}{n_j}\mathbb{1}_{n_j}$ is due to the isomorphism between the affine space of local Hermitian operators, with trace one, acting on $\mathcal{H}_j$ and $\mathfrak{su}^*(\mathcal{H}_j)$, such that $\rho^{(j)}$ be uniquely decomposed as $\rho^{(j)}:= \frac{1}{\text{dim}(\mathcal{H}_j)}+ \text{i}\eta^{(j)}$, where $\text{i} \eta^{(j)} \in \mathfrak{su}^*(\mathcal{H}_j)$ by the Killing-Cartan bilinear form \cite[Section 5.4.1]{chruscinski2004}. Therefore, 
\[
dJ_X(Y_M)(p) = d\langle \textbf{J}(p), X \rangle (Y_M) (p) = i_{X_M}\omega = \omega_p(X_M,Y_M) = - \frac{\text{i}}{2} \text{Tr}(\rho_{\psi}[X,Y]), 
\]
where $X \in \mathfrak{k}$ and for every $Y \in \mathfrak{u}(\mathcal{H})$ with
\[
X_M(p) \equiv \phi_X(p) \in \mathfrak{k}.p := \left\{ \phi_X(p) \in T_pM : \, \phi_X(p) = - \text{i} \, [\eta,p], \eta \in \mathfrak{k}^* \right\},
\]
\[
Y_M(p) \in T_pM := \left\{ Y_M(p) : Y_M(p) = - \text{i} \, [A,p], \, A \in \mathfrak{u}^*(\mathcal{H}) \right\},
\]
with $- \text{i} \eta = X \in \mathfrak{k}$ and $- \text{i}A = Y \in \mathfrak{u}(\mathcal{H})$. 
\section{Singular Reduction of Local Unitary Action} \label{singular-reduction-of-local-unitary-action}

Now consider the projective Hilbert space of tripartite pure states of qubits, as distinguishable particles, i.e. $\mathcal{P}(\mathcal{H}) = \pi(\mathbb{C}^2 \otimes \mathbb{C}^2 \otimes \mathbb{C}^2) \cong \mathbb{C}P(7)$, which is acted upon by the local unitary group $K = SU(2)^{\times_3}$. As it was pointed out previously, $K$ is a compact Lie subgroup of the natural unitary group $U(\mathcal{H})$. The corresponding Lie algebra of $K$ is $\mathfrak{k} = \mathfrak{su}(\mathcal{H}_1) \oplus \mathfrak{su}(\mathcal{H}_2) \oplus \mathfrak{su}(\mathcal{H}_3)$, where $\mathcal{H}_i \cong \mathbb{C}^2$, for $i = 1,2,3$, and is spanned by the matrices of the form $X_1 \otimes \mathbb{1}_2 \otimes \mathbb{1}_2 + \mathbb{1}_2 \otimes X_2 \otimes \mathbb{1}_2 + \mathbb{1}_2 \otimes \mathbb{1}_2 \otimes X_3$, where $X_j \in \mathfrak{su}(2)$ are traceless anti-Hermitian matrices, for $j=1,2,3$. 

Following the notation in \cite{sawicki2011a,sawicki2011b}, any state $\psi \in \mathcal{H}$ can be written as 
\begin{equation}
\psi = \sum_{i_1,i_2,i_3 =0}^{1}{C_{i_1i_2i_3} \, e_{i_1} \otimes e_{i_2} \otimes e_{i_3}},
\label{tripartite-pure-state-eq}
\end{equation}
where $\left\{ e_{i_k} \right\}$ are orthonormal bases for the Hilbert spaces $\mathcal{H}_k$, for $k=1,2,3$. Therefore, the moment map $\textbf{J}:\mathcal{P}(\mathcal{H}) = M \rightarrow \mathfrak{k}^*, p \equiv \rho_{\psi} = \ket{\psi} \bra{\psi}/ \braket{\psi}{\psi} \mapsto \textbf{J}(p)$ can be written as
\begin{equation}
\textbf{J}(p) := (\rho^{(1)}-\frac{1}{2} \mathbb{1}_2) \oplus (\rho^{(2)}-\frac{1}{2} \mathbb{1}_2) \oplus (\rho^{(3)}-\frac{1}{2} \mathbb{1}_2) \in \mathfrak{k}^* = \mathfrak{su}^*(2) \oplus \mathfrak{su}^*(2) \oplus \mathfrak{su}^*(2),
\label{moment-map-3-qubits-eq}
\end{equation}
where $(\rho^{(k)})_{mn} = \sum_{i_1,i_2 = 0}^{1}{\bar{C}_{i_1,\hat{m},i_2} C_{i_1,\hat{n},i_2}}$, and the sum is over all pair indices except $\hat{m}$ and $\hat{n}$ at the $k$th place. Hence, one can write the associated Hamiltonian function $J_X(p)$ to $X \in \mathfrak{k}$ as
\begin{equation}
J_X(p) = \frac{\text{i}}{2} \text{Tr}(X \rho_{\psi}) = \frac{\text{i}}{2} \sum_{k=1}^{3}{\sum_{i_k,j_k=0}^{1}{(\rho^{(k)})_{i_k,j_k} \braket{e_{i_k}}{X_k e_{j_k}}}},
\label{hamiltonian-functions-eq}
\end{equation}
where $X_k \in \mathfrak{su}(2)$, and can be interpreted as the summation of the expectation values of the locally defined Hermitian operators $\text{i}X_k \in \mathfrak{su}^*(2)$ on each laboratory.

Recall that the Lie group $K$ acts on its Lie algebra $\mathfrak{k}$ by adjoint action, given by $\text{Ad}: K \times \mathfrak{k} \rightarrow K, (g,X) \mapsto \text{Ad}_g X = g X g^{-1}$. The corresponding coadjoint action is defined by $\langle \text{Ad}^*_g \xi, X \rangle = \langle \xi, \text{Ad}_{g^{-1}}X \rangle = \langle \xi , g^{-1} X g \rangle$,  for $\xi \in \mathfrak{k}^*$, where $\langle \, . \, , \, . \, \rangle$ represents the natural pairing between $\mathfrak{k}$ and $\mathfrak{k}^*$. Therefore, the resulting orbit $K . \xi = \left\{ \text{Ad}^*_g \xi : g \in K \right\}$ is called the coadjoint orbit. It is well-known \cite{bott1979,kirillov2004} that each coadjoint orbit $K.\xi$ intersects the dual of the Cartan subalgebra, i.e. the maximal commutative subalgebra $\mathfrak{t}^*$ of $\mathfrak{k}^*$, in accordance to the action of the Weyl group $W = N(T)/T$, where $N(T)$ is the normalizer of the maximal torus $T$ of $K$. In fact, the Cartan subalgebra $\mathfrak{t}$ is the Lie algebra of the maximal torus $T$ and $\mathfrak{t} \cong \mathfrak{t}^*$. Hence, modulo the action of the Weyl group $W$ on the Cartan subalgebra $\mathfrak{t}^*$, each coadjoint orbit $K. \xi$ intersects the corresponding positive Weyl chamber $\mathfrak{t}^*_+ \cong \mathfrak{k}^*/K$ only once. In fact, $\mathfrak{t}^*_+$ parametrizes the set of coadjoint orbits in $\mathfrak{k}^*$ and the isotropy subgroup $K_{\xi}$ for the point $\xi \in \mathfrak{t}^*_+$ depends only on the open face of $\mathfrak{t}^*_+$ containing $\xi$ and $K_{\xi} = T$, for $\xi \in \text{int}(\mathfrak{t}^*_+)$ \cite{meinrenken1999}.

In our particular case of interest, the Lie group $SU(2)$ consists of $2 \times 2$ special unitary matrices and the Lie algebra $\mathfrak{su}(2)$ is the space of traceless, anti-Hermitian matrices. The maximal torus $T$ is the subspace of diagonal special unitary matrices and the associated Cartan subalgebra $\mathfrak{t}$ contains traceless, diagonal anti-Hermitian matrices. So, the Weyl group $W$ is the symmetric group $S_2$ acting on the Cartan subalgebra $\mathfrak{t}$ by permuting diagonal elements. Therefore, the positive Weyl chamber $\mathfrak{t}^*_+$ consists of traceless, diagonal Hermitian matrices, such that the diagonal elements are ordered non-increasingly. Hence, the interior of the positive Weyl chamber $\text{int}(\mathfrak{t}^*_+)$ consists of those $\lambda \in \mathfrak{t}^*_+$, such that all their eigenvalues are distinct.

To every compact and connected Hamiltonian $K$-manifold $(M,\omega,K,\textbf{J})$, with the moment map $\textbf{J}:M \rightarrow \mathfrak{k}^*$, is associated a \textit{convex} polytope $\Delta := \textbf{J}(M) \cap \mathfrak{t}^*_+$, called the moment (or Kirwan) polytope \cite{guillemin1982,kirwan1984}. The composite \textit{invariant moment map} $\textbf{J}^{\prime}: M \rightarrow \mathfrak{k}^* \rightarrow \mathfrak{t}^*_+, p \mapsto \textbf{J}^{\prime}(p) = \textbf{J}(K.p) \cap \mathfrak{t}^*_+$ is an open map onto its image \cite{knop2002}, such that all its fibers are connected \cite{kirwan1984}. Hence, the points $\xi \in \Delta = \textbf{J}^{\prime}(M) \subset \mathfrak{t}^*_+$ are sufficient to find the symplectic quotients $M_{\xi}$ and their strata $M_{\xi}^{(H)}$, where $\xi \in \Delta$ and $H < K$.

Recalling the singular symplectic method \cite{sjamaar1991,bates1997,ortega2004}, associated to each singular value $\xi = \textbf{J}(p) \in \mathfrak{k}^*$ is a symplectic reduced space $M_{\xi} = \textbf{J}^{-1}(\xi)/K_{\xi}$, which is a stratified symplectic space, with the isotropy subgroup $K_{\xi} = \{ g \in K : \, \text{Ad}_g^{*} \xi = \xi \}$, namely,
\[
M_{\xi} := \textbf{J}^{-1}(\xi)/K_{\xi} = \bigcup_{H < K} M_{\xi}^{(H)},
\]
where each stratum $M_{\xi}^{(H)}$ denotes the equivalence class of all the points $p \in \textbf{J}^{-1}(\xi)$ whose isotropy subgroups $K_p$ are conjugate to $H$. Moreover, there exists a unique principal stratum $M_{\xi}^{(\text{prin})}$, which is open, dense and connected in the reduced space $M_{\xi}$. In the next subsection \ref{principal-stratum-of-symplectic-quotient}, the main results of this paper, namely the local normal form, the symplectic structure and the components of the reduced dynamics on the principal stratum of the symplectic reduced space are studied in further details.

\subsection{Principal Stratum of Symplectic Quotient} \label{principal-stratum-of-symplectic-quotient}

In \cite{sawicki2012b}, by using the fact that the principal orbit type stratum $M_{(\text{prin})}$ is a connected, open and dense manifold of the orbit type stratification of $M$, and is of maximum dimension, since the isotropy subgroups $K_p$ of the generic points $p \in M$ of the projective Hilbert space of tripartite pure states are discrete \cite{carteret2000}, it is discussed that the image of the composite invariant moment map $\textbf{J}^{\prime}(M_{(\text{prin})})$ contains the relative interior of the moment polytope $\text{int}(\Delta) = \textbf{J}(M_{(\text{prin})}) \cap \text{int}(\mathfrak{t}^*_+)$. Also it is shown that the principal stratum $M^{(\text{prin})}_{\xi} = (\textbf{J}^{-1}(\xi) \cap M_{(\text{prin})})/K_{\xi}$ of the symplectic quotient $M_{\xi} = \textbf{J}^{-1}(\xi)/K_{\xi}$, for all $\xi \in \text{int}(\Delta)$, is two dimensional.

Recalling the symplectic stratification theorem in section \ref{review-of-symplectic-singular-reduction}, $M_{\xi}$, for $\xi \in \text{int}(\Delta) \subset \mathfrak{t}^*_+$, is a stratified symplectic space, with $(M_{\xi}^{(H)}, \omega^{(H)}_{\xi})$ as the symplectic strata, for all $H < K$. To find the reduced symplectic structure $\omega^{(\text{prin})}_{\xi}$ on the principal stratum of $M^{(\text{prin})}_{\xi}$, for the principal isotropy subgroup $H_{\text{prin}}$, we have to note that $\textbf{J}^{-1}(\xi) \cap M_{(\text{prin})}$ is a submanifold of $M$, and also
\[
(i^{(\text{prin})}_{\xi})^* \omega = (\pi^{(\text{prin})}_{\xi})^* \omega^{(\text{prin})}_{\xi},
\]
i.e.
\begin{equation}
\omega_p(Z_M(p),Z_M^{\prime}(p)) = \omega^{(\text{prin})}_{\xi}(T\pi^{(\text{prin})}_{\xi}(Z_M(p)), T\pi^{(\text{prin})}_{\xi}(Z^{\prime}_M(p))),
\label{symplectic-structure-of-principal-stratum-eq}
\end{equation}
at $p \in \textbf{J}^{-1}(\xi) \cap M_{(\text{prin})}$, where $Z_M(p),Z^{\prime}_M(p) \in T_p(\textbf{J}^{-1}(\xi) \cap M_{(\text{prin})})$, and
\[
T\pi^{(\text{prin})}_{\xi}: T_p(\textbf{J}^{-1}(\xi) \cap M_{(\text{prin})}) \rightarrow T_xM^{(\text{prin})}_{\xi}, \quad x = \pi^{(\text{prin})}_{\xi}(p) \in M^{(\text{prin})}_{\xi}.
\]
Moreover, as it is proposed in \cite{sawicki2012b}, we have $T_p(\textbf{J}^{-1}(\xi) \cap M_{(\text{prin})}) = (\mathfrak{k}.p)^{\omega}$, where $(\,)^{\omega}$ represents $\omega$-orthogonality, since the map $d_p\textbf{J}: T_pM \rightarrow \mathfrak{k}^*$ is surjective at $p \in M_{(\text{prin})}$ with discrete isotropy subgroups $K_p = H_{\text{prin}}$. Therefore \cite[Section 4.3]{abraham1987},
\begin{equation}
T_xM^{(\text{prin})}_{\xi} \cong T_p(\textbf{J}^{-1}(\xi) \cap M_{(\text{prin})})/(\mathfrak{k}_{\xi}.p) \equiv V_x,
\label{tangent-space-to-principal-stratum-eq}
\end{equation}
where $\mathfrak{k}_{\xi}.p = \{ X_M(p) \equiv \phi_X(p) \in \mathfrak{k}.p : \, \phi_X(p) = -\text{i} \, [\eta,p], \, \eta \in \mathfrak{k}^*_{\xi} \cong \mathfrak{t}^* \}$, for $-\text{i}\eta \equiv X \in \mathfrak{k}_{\xi} \cong \mathfrak{t}$ and $\xi \in \text{int}(\Delta)$. In fact, the space $\mathfrak{k}_{\xi}.p$ is the degeneracy space introduced in \cite{sawicki2011a}, for which the dimension represents a measure for quantum entanglement of pure states. Locally, the subspace $V_x \equiv (\mathfrak{k}.p)^{\omega}/(\mathfrak{k}_{\xi}.p) \cong (\mathfrak{k}.p)^{\omega}/((\mathfrak{k}.p)^{\omega} \cap (\mathfrak{k}.p))$ is the symplectic subspace of $(\mathfrak{k}.p)^{\omega}$ in the Witt-Artin decomposition of $T_pM$ for the Hamiltonian action of the compact Lie group $K$ on $(M,\omega)$ \cite[Section 7.1]{ortega2004}. It is also called the \textit{symplectic normal space} in the Marle-Guillemin-Sternberg local normal form \cite{guillemin1984,marle1985}, as a local model for a symplectic manifold $M$ equipped with a Hamiltonian action of a compact Lie group $K$ around any orbit $K.p$ in the fiber $\textbf{J}^{-1}(\xi)$, namely
\[
Y = K \times_{K_p} \left( \left( \mathfrak{k}_{\xi} / \mathfrak{k}_p \right)^* \times V_x \right),
\]
whose elements are equivalence classes $[g,\rho,v]$, by the free action of the isotropy subgroup $K_p \equiv H$ on $K \times \left( \left( \mathfrak{k}_{\xi} / \mathfrak{k}_p \right)^* \times V_x \right)$, given by $h \, . \, (g,\rho,v) = (gh^{-1}, \text{Ad}^*_{h^{-1}} \rho,h.v)$. The symplectic normal space $V_x$ is defined as in the Eq. \eqref{tangent-space-to-principal-stratum-eq}. There exists a $2$-from $\omega_Y$ on $Y$, which is symplectic near $[e,0,0]$ and so the orbit $K.p$ can be considered as the zero section in the normal bundle $Y$. Also the Lie group $K$-action on $Y$, which is given by $g^{\prime} \, . \, [g,\rho,v] = [g^{\prime}g,\rho,v]$ is a Hamiltonian action and therefore is equipped with a moment map $\textbf{J}_Y: Y \rightarrow \mathfrak{k}^*$, given by
\[
\textbf{J}_Y \left( [g,\rho,v] \right) = \text{Ad}_{g^{-1}}^*(\xi + \rho + \textbf{J}_V(v)),
\]
where the moment map $\textbf{J}_V : V_x \rightarrow \mathfrak{h}^*$ for the linear $H$-action on the symplectic normal space $V_x$ is quadratic homogeneous and is described below. Therefore, the Hamiltonian $K$-manifold $(M,\omega,K,\textbf{J})$ is locally modeled by $(Y,\omega_Y,K,\textbf{J}_Y)$ around every orbit $K.p$ at $p$.

Recall that the closed subgroups of the Lie group $SU(2)$ are as follows: the group $SU(2)$ itself, the maximal torus $U(1)$, the normalizer in $SU(2)$ of the maximal torus $U(1)$ and a collection of finite subgroups. Therefore, the principal isotropy subgroup $H_{\text{prin}}$ belongs to the set of finite subgroups of the Lie group $K = SU(2)^{\times_3}$, collectively denoted by $\Gamma$. Hence, the local normal form $Y$ around the orbit $K.p$, where $p \in \textbf{J}^{-1}(\xi) \cap M_{(\text{prin})}$, is given by
\begin{eqnarray}
Y & = & K \times_{H_{\text{prin}}} \left( \mathfrak{k}^*_{\xi} \times (\mathfrak{k}.p)^{\omega}/(\mathfrak{k}_{\xi}.p) \right) \nonumber \\
& = & \left\{ \left. [g,\rho,v] \right| g \in K, \, \rho \in \mathfrak{k}^*_{\xi} \cong \mathfrak{t}^*, \, v \in (\mathfrak{k}.p)^{\omega}/(\mathfrak{k}_{\xi}.p) , \right. \nonumber \\
& & \left. [g,\rho,v] = [gh^{-1},\text{Ad}^*_{h^{-1}} \rho, h. v], \, \forall h \in H_{\text{prin}} \right\},
\label{local-normal-form-eq}
\end{eqnarray}
with $\textbf{J}_Y([g,\rho,v]) = \text{Ad}^*_{g^{-1}}(\xi + \rho)$, since for discrete isotropy subgroup $K_p \equiv H_{\text{prin}}$, the moment map $\textbf{J}_V$ is the zero map and $\textbf{J}_V(v) = 0$, for every $v \in V_x$, and the isotropy subalgebra $\mathfrak{k}^*_p$ is trivial. 

In fact the fixed point set $V_x^H$ of the symplectic normal space $V_x$ is locally isomorphic to the tangent space to the symplectic strata $T_xM^{(H)}_{\xi}$, for $H < K$ and $\xi \in \mathfrak{k}^*$. More precisely, in the local normal form the symplectic normal space $V_x$ is acted upon properly, linearly and in a Hamiltonian fashion by the isotropy subgroup $K_p \equiv H$, and is therefore equipped with the associated moment map $\textbf{J}_V : V_x \rightarrow \mathfrak{h}^*,\, u \mapsto \textbf{J}_V(u)$, defined by $\langle \textbf{J}_V , X \rangle = \frac{1}{2} \omega_V(X.u,u)$, for every $X \in \mathfrak{h}$ and $u \in V_x$,  where $\omega_V$ is the symplectic structure of $V_x$. Hence, $\textbf{J}_V(u) = 0$, for every $u \in V_x^H = \{ u \in V_x : \, h.u = u, \, \forall h \in H \}$. Therefore, $V_x^H \cong V_x$, if $H$ is a discrete isotropy subgroup, as is the case here for $H = H_{\text{prin}}$. 

In other words,
\begin{eqnarray}
T\pi^{(\text{prin})}_{\xi} & : & (\mathfrak{k}.p)^{\omega} \rightarrow V_x \equiv T_xM^{(\text{prin})}_{\xi} \cong (\mathfrak{k}.p)^{\omega}/(\mathfrak{k}_{\xi}.p), \nonumber \\
& & Z_M(p) \mapsto T\pi^{(\text{prin})}_{\xi}(Z_M(p)) \in V_x,
\label{tangent-projection-map-eq}
\end{eqnarray}
where $(\mathfrak{k}.p)^{\omega} = \{Y_M(p) \in T_pM : \, \omega_p(X_M(p),Y_M(p)) = 0, \, \forall X_M(p) \equiv \phi_X(p) \in \mathfrak{k}.p \}$, can be re-written as
\[
(\mathfrak{k}.p)^{\omega} = \left\{ Y_M(p) \in T_pM : \, \omega_p(X_M(p),Y_M(p)) = - \frac{\text{i}}{2} \text{Tr}(\rho_{\psi}[X,Y])= 0, \, \forall X \in \mathfrak{k} \right\},
\]
where $X_M(p) \equiv \phi_X(p) = - \text{i} [\eta,\rho_{\psi}]$, for $\eta \in \mathfrak{k}^*$, i.e. $-\text{i} \eta = X \in \mathfrak{k}$, and $Y_M(p) = - \text{i} [A,\rho_{\psi}]$, for $A \in \mathfrak{u}^*(\mathcal{H})$, i.e. $- \text{i}A = Y \in \mathfrak{u}(\mathcal{H})$. Therefore, if we define
\begin{equation}
V^{\prime}_x := \left\{ A \in \mathfrak{u}^*(\mathcal{H}) : \, \text{Tr}(\rho_{\psi}\,[\eta,A]) = 0, \forall \eta \in \mathfrak{k}^*/\mathfrak{k}^*_{\xi} \cong \mathfrak{k}^*/\mathfrak{t}^* \right\},
\label{vprime-space-eq}
\end{equation}
then, $V_x = \{ Y_M(p) \in T_pM : \, \omega_p(X_M(p),Y_M(p)) = 0, \, \forall X_M(p) \equiv \phi_X(p) \in \mathfrak{k}.p/\mathfrak{t}.p \}$, can be re-written as
\begin{equation}
V_x = \left\{ T\pi^{(\text{prin})}_{\xi}(Y_M(p)) \in T_pM : \, T\pi^{(\text{prin})}_{\xi}(Y_M(p)) = - \text{i} \, [A,\rho_{\psi}], \,  A \in V^{\prime}_x \right\}.
\label{symplectic-subpace-of-witt-artin-decomposition-eq}
\end{equation}

Equivalently, $\omega_p(X_M(p),Y_M(p)) = 0$, if and only if $dJ_X(Y_M)(p)=0$, since 
\[
dJ_X(Y_M)(p) = \omega_p(X_M(p),Y_M(p)),
\]
for $X_M(p) \in \mathfrak{k}.p$ and $Y_M(p) \in T_pM$. Hence, the symplectic structure $\omega^{(\text{prin})}_{\xi}$ on the principal stratum $M^{(\text{prin})}_{\xi}$ is given by
\begin{equation}
\omega^{(\text{prin})}_{\xi}(T\pi^{(\text{prin})}_{\xi}(Z_M(p)), T\pi^{(\text{prin})}_{\xi}(Z^{\prime}_M(p))) =  \frac{\text{i}}{2} \text{Tr}(\rho_{\psi} \, [Z,Z^{\prime}]),
\label{symplectic-structure-of-principal-stratum-eq2}
\end{equation}
where $Z_M(p),Z^{\prime}_M(p) \in (\mathfrak{k}.p)^{\omega}$, and $T\pi^{(\text{prin})}_{\xi}(Z_M(p)), T\pi^{(\text{prin})}_{\xi}(Z^{\prime}_M(p)) \in V_x$, namely $Z,Z^{\prime} \in V^{\prime}_x$. Therefore, in our particular case of interest, if $\{e_{i_j}\}$, with $i_j =0,1$, represent orthonormal bases for the Hilbert spaces $\mathcal{H}_j$, for $j = 1,2,3$, then $\{ e_{i_1} \otimes e_{i_2} \otimes e_{i_3}: \, i_j = 0,1, \, \, j= 1,2,3\}$ represents an orthonormal basis for the Hilbert space $\mathcal{H} \cong \otimes_{j=1}^3 \mathcal{H}_j$. Recalling the Eq. \eqref{tripartite-pure-state-eq}, $V^{\prime}_x$ can be obtained by finding those non-local Hermitian operators $A \in \mathfrak{u}^*(\mathcal{H})$ such that
\begin{eqnarray}
\text{Tr}(\rho_{\psi}\,[\eta,A]) & = & \sum_{i_1,i_2,i_3=0}^{1}{\sum_{j_1,j_2,j_3=0}^{1}{\bar{C}_{i_1i_2i_3} C_{j_1j_2j_3} \braket{e_{i_1} \otimes e_{i_2} \otimes e_{i_3}}{[\eta,A] e_{j_1} \otimes e_{j_2} \otimes e_{j_3}}}} \nonumber \\
& = & 0, 
\label{omega=0-eq}
\end{eqnarray}
for all $\eta \equiv \eta_1 \oplus \eta_2 \oplus \eta_3$, with $\eta_k \in \mathfrak{su}^*(2)/\mathfrak{t}^*(2)$, for $k =1,2,3$. Following \cite{sawicki2011a}, let $E_{ij}$ denotes the matrix with $1$ in the $(i,j)$-position, with $i \neq j$, and zero elsewhere, then the matrices $E_{ij} + E_{ji}$ and $\text{i} \, (E_{ij} - E_{ji})$ form a basis for $\mathfrak{su}^*(2)/\mathfrak{t}^*(2) \cong \mathbb{R}^2$, i.e. spanned by
\[
\left\{ \left( \begin{array}[pos]{cc} 0 & 1 \\ 1 & 0 \end{array}  \right) , \left( \begin{array}[pos]{cc} 0 & - \text{i} \\ \text{i} & 0 \end{array} \right) \right\},
\]
since for each group $SU(2)=S^3$ we have $SU(2) \rightarrow SU(2)/T$, as the Hopf fibration, namely $SU(2)/U(1) \cong S^2 = \mathbb{C}P(1)$.

\subsection{Reduced Dynamics on the Principal Stratum} \label{reduced-dynamics-on-principal-stratum}

The symplectic structure $\omega^{(\text{prin})}_{\xi}$ on the principal stratum $M^{(\text{prin})}_{\xi}$ allows us to define on the space of smooth functions $C^{\infty}(M^{(\text{prin})}_{\xi})$ a Poisson bracket $\{ \, . \, , \, . \, \}_{(\text{prin})}$ as
\begin{equation}
\left\{ f^{(\text{prin})}_{\xi}, g^{(\text{prin})}_{\xi} \right\}_{(\text{prin})}(x) = \omega^{(\text{prin})}_{\xi}(X_f(x),X_g(x)),
\label{poisson-bracket-on-principal-stratum-eq}
\end{equation}
where $X_f, X_g$ are the corresponding Hamiltonian vector fields of the reduced Hamiltonian functions $f^{(\text{prin})}_{\xi},g^{(\text{prin})}_{\xi} \in C^{\infty}(M^{(\text{prin})}_{\xi})$, defined by the Eq. \eqref{reduced-hamiltonian-functions-on-the-strata} as
\[
(\pi^{(\text{prin})}_{\xi})^*f^{(\text{prin})}_{\xi} = (i^{(\text{prin})}_{\xi})^*f,
\]
namely
\begin{equation}
f^{(\text{prin})}_{\xi}(\pi^{(\text{prin})}_{\xi}(p)) = f(p) = \frac{1}{2} \, \text{Tr}(F \rho_{\psi}), \quad p \in \textbf{J}^{-1}(\xi) \cap M_{(\text{prin})},
\label{reduced-hamiltonian-functions-eq}
\end{equation}
where $f(p)$s are the $K$-invariant smooth functions, for all $-\text{i}F \in \mathfrak{u}(\mathcal{H})$, i.e. $f(p) \in C^{\infty}(M)^K$, for all $\rho_{\psi} \equiv p \in M$, since the Hermitian structure $\langle \, , \, \rangle$ on the Hilbert space $\mathcal{H}$ is $K$-invariant. In other words,
\begin{equation}
(\pi^{(\text{prin})}_{\xi})^* \left\{ f^{(\text{prin})}_{\xi} , g^{(\text{prin})}_{\xi} \right\}_{(\text{prin})}(x) = (i^{(\text{prin})}_{\xi})^*\left\{ f, g \right\}(p) = \left\{ f , g \right\}(p), 
\label{reduced-hamiltonian-eq}
\end{equation}
where $p \in \textbf{J}^{-1}(\xi) \cap M_{(\text{prin})}$ and $f,g \in C^{\infty}(M)^K$ denote the corresponding $K$-invariant smooth Hamiltonian functions. Recalling the Eq. \eqref{symplectic-structure-of-principal-stratum-eq2}, the induced Hamiltonian vectors on $T_xM^{(\text{prin})}_{\xi}$ is given by
\begin{equation}
X_f(x) = X_f(\pi^{(\text{prin})}_{\xi}(p)) \equiv T\pi^{(\text{prin})}_{\xi}(X^{(f)}_M(p)) \in V_x,
\label{induced-hamiltonian-vectors-eq}
\end{equation}
such that $X^{(f)}_M(p) \in (\mathfrak{k}.p)^{\omega} \subset T_pM$, for $F \in V^{\prime}_x$. Hence, the expectation values $f^{(\text{prin})}_{\xi}(x) = \frac{1}{2} \, \text{Tr}(F \rho_{\psi})$, such that $\rho_{\psi} = p \in \textbf{J}^{-1}(\xi) \cap M_{(\text{prin})}$ and for the Hermitian operators $F \in V^{\prime}_x$ are the smooth reduced Hamiltonian functions generating the reduced Hamiltonian dynamics on the principal stratum $M^{(\text{prin})}_{\xi}$, namely
\begin{equation}
df^{(\text{prin})}_{\xi}(X_g)(x) = i_{X_g}\omega^{(\text{prin})}_{\xi} = \omega^{(\text{prin})}_{\xi}(X_f(x),X_g(x)) = \frac{\text{i}}{2} \text{Tr}(\rho_{\psi}[F,G]),
\label{reduced-hamiltonian-vector-fields-eq2}
\end{equation}
for every $G \in V^{\prime}_x$ and with the induced Hamiltonian vector fields $X_f(x) = - \text{i} \,[F,\rho_{\psi}] \in V_x$ and $X_g(x) = - \text{i} \, [G,\rho_{\psi}] \in V_x$. The Poisson bracket \eqref{poisson-bracket-on-principal-stratum-eq} can help us to determine the Hamiltonian flows on the principal stratum $M^{(\text{prin})}_{\xi}$. Let $\varphi_t^{(\text{prin})}(x)$ and $\varphi_t(p)$ denote the Hamiltonian flows of $f^{(\text{prin})}_{\xi} \in C^{\infty}(M^{(\text{prin})}_{\xi})$ and $f(p) \in C^{\infty}(M)^K$, such that $f^{(\text{prin})}_{\xi}(\varphi_t^{(\text{prin})}(x)) = f(\varphi_t(p))$, for $p \in \textbf{J}^{-1}(\xi) \cap M_{(\text{prin})}$ and $x = \pi^{(\text{prin})}_{\xi}(p)$. Then,
\begin{eqnarray}
\frac{dg^{(\text{prin})}_{\xi}}{dt}(\varphi_t^{(\text{prin})}(x)) & = & \frac{dg}{dt}(\varphi_t(p)) = \left\{ g,f \right\}(\varphi_t(p)) \nonumber \\
& = & \left\{ g^{(\text{prin})}_{\xi}, f^{(\text{prin})}_{\xi} \right\}_{(\text{prin})}(\varphi_t^{(\text{prin})}(x)),
\label{reduced-flow-eq}
\end{eqnarray}
for every $g^{(\text{prin})}_{\xi} \in C^{\infty}(M^{(\text{prin})}_{\xi})$, such that $g^{(\text{prin})}_{\xi}(x) = g(p)$. Therefore, from the Eqs. \eqref{poisson-bracket-on-principal-stratum-eq} and \eqref{reduced-flow-eq}, it is implied that
\begin{equation}
df^{(\text{prin})}_{\xi}(x)(.) = \omega^{(\text{prin})}_{\xi}(X_f(x), \, . \,),
\label{reduced-hamiltonian-vector-fields-eq}
\end{equation}
where $X_f(p) \in V_x$, for all $f^{(\text{prin})}_{\xi}(x) = \frac{1}{2} \text{Tr}(F \rho_{\psi})$, such that $F \in V^{\prime}_x$, $\xi \in \text{int}(\Delta)$ and $p \equiv \rho_{\psi} \in \textbf{J}^{-1}(\xi) \cap M_{(\text{prin})}$.

In \cite{sjamaar1991}, it is shown that the connected components of the strata are symplectic leaves of the quotient $M_{\xi}$, i.e. let $x_1, x_2$ be two points in the connected component of a stratum in $M_{\xi}$, then there exists a piecewise smooth path joining $x_1$ to $x_2$ consisting a finite number of Hamiltonian trajectories of smooth functions in $C^{\infty}(M_{\xi})$, since their Hamiltonian flows preserve the stratification and also the restriction of their flows to a stratum equals to the Hamiltonian flows of the reduced Hamiltonians. 

Put it another way, as in \cite{cushman2001}, a continuous curve $\varphi_t^{(\text{prin})}: [t_1,t_2] \rightarrow M^{(\text{prin})}_{\xi}$ is a piecewise integral curve of Hamiltonian vector fields of smooth reduced Hamiltonians, if the interval $[t_1,t_2]$ can be partitioned into finite number of sub-intervals $[t_j,t_{j+1}]$, for $j=1, \cdots ,k$, such that the restriction of the flow $\varphi_t^{(\text{prin})}$ to the sub-interval $[t_j,t_{j+1}]$, i.e. $\varphi_{t_j}^{(\text{prin})}: [t_j,t_{j+1}] \rightarrow M^{(\text{prin})}_{\xi}$, is the integral curve of the Hamiltonian vector field $X_f$ of a reduced Hamiltonian $f^{(\text{prin})}_{\xi}$, namely the solution of the Eq. \eqref{reduced-flow-eq}, for every $t \in [t_j,t_{j+1}]$ and every $g^{(\text{prin})}_{\xi} \in C^{\infty}(M^{(\text{prin})}_{\xi})$. Then every two points $x_1,x_2 \in M^{(\text{prin})}_{\xi}$ can be joined by a piecewise integral curves of Hamiltonian vector fields.

Quantum mechanically, given a (shifted) spectra of the single-particle reduced density matrices, such that the eigenvalues of each particle are distinct and are ordered non-increasingly, then the associated global pure state, represented by a generic point $p$ with one entanglement type, can be transformed to another pure state (generic point) with another entanglement type and the same value of the moment map spectra, under the flow $\varphi_t^{(\text{prin})}$, namely by a finite sequence of integral curves of induced Hamiltonian vector fields of smooth reduced Hamiltonian functions.

Moreover, the symplectic normal space $V_x$, or better $V^{\prime}_x$ in the Eq. \eqref{vprime-space-eq}, represents the space of non-local time-independent quantum control Hermitian operators (Hamiltonians), which can induce unitary entanglement dissipation \cite{solomon2012} for the generic points of a composite quantum system containing three qubits. However, this is not a true dissipation process, since the reduced flow $\varphi_t^{(\text{prin})}$ in the principal stratum $M^{(\text{prin})}_{\xi}$ consists of a sequence of one-parameter family of local diffeomorphisms corresponding to the induced Hamiltonian vector fields and so a reversible process. Of course one has to note that the reduced dynamics and so the unitary entanglement dissipation only occurs on $M^{(\text{prin})}_{\xi}$, for instance the entanglement type of the separable $p_s$ or bi-separable $p_{b_k}$ states, for $k=1,2,3$, can not be changed unitarily as discussed above, since $\textbf{J}^{\prime}(p_s)$ and $\textbf{J}^{\prime}(p_{b_k})$ are not included in the $\text{int}(\Delta)$ and so $p_s,p_{b_k} \notin \textbf{J}^{-1}(\xi) \cap M_{(\text{prin})}$. The symplectic reduced space for other values of the moment map not included in the interior of the moment polytope $\Delta$ will be further discussed in the subsequent section \ref{dynamics-on-the-other-strata}.

\subsection{Dynamics on the Other Strata} \label{dynamics-on-the-other-strata}

By definition of a stratified symplectic space, for instance $M_{\xi}$, for a fixed $\xi \in \text{int}(\Delta)$ together with the algebra $C^{\infty}(M_{\xi})$ of smooth functions on $M_{\xi}$, the following conditions are satisfied: each stratum $S$ is a symplectic manifold; $C^{\infty}(M_{\xi})$ is a Poisson algebra and the embedding $S \hookrightarrow M_{\xi}$ is Poisson \cite{lerman1994}. From the last condition, it is implied that the symplectic structure on the open dense stratum, i.e. the principal stratum $M_{\xi}^{(\text{prin})}$, determines the symplectic structures on all other lower dimensional strata $M_{\xi}^{(H)}$, for $H_{(\text{prin})} \neq H < K$, and so the Poisson structure on the whole symplectic quotient $M_{\xi}$.

If, for a given $\xi \in \text{int}(\Delta)$, the principal stratum $(M^{(\text{prin})}_{\xi}, \omega^{(\text{prin})}_{\xi})$ is a two dimensional symplectic manifold, then all other lower dimensional strata $M^{(H)}_{\xi}$, for $H_{(\text{prin})} \neq H < K$, would be zero dimensional, i.e. isolated points, since all the strata $M^{(H)}_{\xi}$ are symplectic manifolds. Therefore, they are the fixed points of all Hamiltonian vector fields $X_M^{(f)}$ on $M$, for every $f_{\xi}^{(H)}(\pi_{\xi}^{(H)}(p)) = f(i_{\xi}^{(H)}(p))$, such that $f \in C^{\infty}(M)^K$ and $f_{\xi}^{(H)} \in C^{\infty}(M_{\xi}^{(H)})$, and $\pi_{\xi}^{(H)}: \textbf{J}^{-1}(\xi) \cap M_{(H)} \rightarrow M_{\xi}^{(H)}$ and $i_{\xi}^{(H)}:\textbf{J}^{-1}(\xi) \cap M_{(H)} \hookrightarrow M$. In other words, they are \textit{relative equilibria} in the projective Hilbert manifold $M$ \cite{sjamaar1991}. Recall that a point $p \in M$ is called a relative equilibrium, if and only if the integral curves of a Hamiltonian vector field $X_M^{(h)}$, for $h \in C^{\infty}(M)^K$, is contained in the orbit $K.p$, so every point in the orbit $K.p$ is also a relative equilibrium. The situation is the same for other points $\xi \neq 0$ on the boundary of the moment polytope $\Delta$, since as it is shown in \cite{sawicki2012b}, their symplectic reduced spaces $M_{\xi}$ are zero dimensional and so they represent relative equilibria in the original manifold $M=\mathcal{P}(\mathcal{H})$ too. 
\section{Conclusions} \label{conclusions}

In this paper, the singular symplectic reduction procedure is applied to the projective Hilbert space of tripartite pure quantum states, under the local unitary group action, for a system consisting of three qubits. Given the (shifted) spectra of the single-particle reduced density matrices, as the components of the associated moment map, such that the eigenvalues of each particle are distinct and are ordered non-increasingly, the symplectic structure on the principal stratum is obtained and it is shown that the Eq. \eqref{omega=0-eq} provides us with a criterion from which the elements of the local normal model on the principal stratum of the symplectic quotient $M_{\xi}$ can be constructed up to the action of the principal isotropy subgroup.

Moreover, from the symplectic structure of the open, dense and connected principal stratum, the induced Hamiltonian vector fields, the reduced smooth Hamiltonian functions and their corresponding reduced Hamiltonian flows are investigated on the principal stratum. Furthermore, it is discussed that for a given spectra of the single-particle reduced density matrices, other lower dimensional strata are isolated points and so they are the fixed points of every reduced Hamiltonian flow, which are known as the relative equilibria in the original manifold $M$. 

From physical point of view, the reduced Hamiltonian flow on the principal stratum, which contains a finite sequence of the integral curves of the induced Hamiltonian vector fields, provides a reversible unitary entanglement dissipation for a composite quantum system containing three qubits. Each reduced Hamiltonian function can be obtained locally from the space of time-independent quantum control Hermitian operators. Finally, while the original projective Hilbert space is a K\"ahler manifold, the metric structure on the symplectic reduced space, and in particular on the principal stratum, as well as the exact computation of the symplectic normal space, will be discussed elsewhere.

\section*{Acknowledgements} \label{acknl}
This work is partially supported by the Malaysian Ministry Of Higher Education (MOHE), Fundamental Research Grant Scheme (FRGS) with Vote No. 5523927.

\renewcommand*{\refname}{Bibliography} 
\bibliographystyle{notunsrtnat}
\bibliography{Bib_symplectic-quotient-of-3qubits_arxiv}

\vspace{.2in}
\noindent
Laboratory of Computational Sciences and Mathematical Physics, \\
Institute for Mathematical Research, Universiti Putra Malaysia, \\
43400 UPM Serdang, Selangor, Malaysia \\
E-mail: \href{mailto:saeid.molladavoudi@gmail.com}{\texttt{saeid.molladavoudi@gmail.com}}

\end{document}